\documentclass[sigconf]{acmart}

\AtBeginDocument{%
  }

\usepackage[utf8]{inputenc}
\usepackage{tabularx}
\usepackage{booktabs}
\usepackage{soul}
\usepackage{xcolor}
\usepackage{xspace}
\usepackage{graphicx}
\usepackage[caption=false]{subfig}  
\usepackage[export]{adjustbox}
\usepackage{amsmath}
\usepackage{textcomp}
\usepackage{listings}
\usepackage{cleveref}
\usepackage{algorithm}
\usepackage[noend]{algpseudocode}
\usepackage{balance}
\usepackage{enumitem}
\usepackage{multirow}
\usepackage{colortbl}
\usepackage{balance}
\usepackage{threeparttable}
\usepackage{multicol}
\usepackage{tcolorbox}

\definecolor{codegreen}{rgb}{0,0.6,0}
\definecolor{codegray}{rgb}{0.5,0.5,0.5}
\definecolor{codepurple}{rgb}{0.58,0,0.82}
\definecolor{backcolour}{rgb}{0.95,0.95,0.92}
\definecolor{gray}{gray}{0.9}
\definecolor{APA_stats}{RGB}{100, 100, 120}

\newcounter{observation}

\setcopyright{acmcopyright}
\copyrightyear{2023}
\acmYear{2023}
\acmDOI{XXXXXXX.XXXXXXX}

\acmConference[ICSE 2024]{46th International Conference on Software Engineering}{April 2024}{Lisbon, Portugal}
\acmPrice{15.00}
\acmISBN{978-1-4503-XXXX-X/18/06}

\begin{document}


\title{In-IDE Human-AI Experience in the Era of Large Language Models; A Literature Review}



\author{Agnia Sergeyuk}
\email{agnia.sergeyuk@jetbrains.com}
\affiliation{%
    \institution{JetBrains Research}
    \city{Belgrade}
    \country{Serbia}
}

\author{Sergey Titov}
\email{sergey.titov@jetbrains.com}
\affiliation{%
  \institution{JetBrains Research}
  \city{Paphos}
  \country{Cyprus}
}

\author{Maliheh Izadi}
\email{m.izadi@tudelft.nl}
\affiliation{%
  \institution{Delft University of Technology}
  \city{Delft}
  \country{Netherlands}
}

\renewcommand{\shortauthors}{Sergeyuk et al.}

\begin{abstract}

    Integrated Development Environments (IDEs) have become central to modern software development, especially with the integration of Artificial Intelligence (AI) to enhance programming efficiency and decision-making. The study of in-IDE Human-AI Experience is critical in understanding how these AI tools are transforming the software development process, impacting programmer productivity, and influencing code quality. 
    
    We conducted a literature review to study the current state of in-IDE Human-AI Experience research, bridging a gap in understanding the nuanced interactions between programmers and AI assistants within IDEs. By analyzing 36 selected papers, our study illustrates three primary research branches: \textbf{Design}, \textbf{Impact}, and \textbf{Quality} of Interaction. 
    
    The trends, challenges, and opportunities identified in this paper emphasize the evolving landscape of software development and inform future directions for research, and development in this dynamic field. Specifically, we invite the community to investigate three aspects of these interactions: designing \textbf{task-specific user interface}, building \textbf{trust}, and improving \textbf{readability}.
\end{abstract}





\maketitle

\section{INTRODUCTION}

In the ever-evolving landscape of technology, Human-Computer Interaction (HCI) stands as a key field, encompassing the study of design, implementation, and usage of computer technology~\cite{rapp2023human}. This dynamic discipline, bridging hardware and software, delves deep into understanding and optimizing the interaction between humans and digital systems. HCI not only investigates how users experience these technologies but also aims to enhance this interaction to be more intuitive, efficient, and user-centric. 



However, the landscape of HCI has been significantly transformed with the advent and evolution of Large Language Models (LLMs) and advanced Artificial Intelligence (AI) assistance tools. This technological progression has shifted the focus from conventional HCI paradigms to a more nuanced and interactive domain, known as Human-AI Experience (HAX)~\cite{amershi2019guidelines}. HAX represents a deeper integration of AI within user interactions, where AI is not just a tool but a collaborative partner, reshaping the way users engage with and benefit from technological systems. This shift underscores the growing importance of understanding and designing for interactions that are increasingly AI-driven, marking a pivotal change in the field of Human-Computer Interaction.

With this change, Integrated Development Environments (IDEs) as the primary platform for Developer-AI interactions, are transforming beyond mere code creation tools into sophisticated ecosystems where AI's role extends from assisting in coding to offering intelligent insights and recommendations. 


As a result, the focus of in-IDE HCI studies, initially concentrated on optimizing day-to-day programming activities from an HCI perspective, is now shifting towards in-IDE HAX, a more integrated and interactive approach. This transition reflects a growing interest in not just improving functional aspects of the developer’s experience, but also in understanding and enhancing the collaborative dynamics between the developer and AI within the IDE.

As the field of in-IDE HAX evolves, there is a growing need for centralized knowledge and collaboration within the community. A collaborative approach can enhance efficiency, minimize redundant efforts, and encourage inspiration among researchers, contributing to the advancement of the field. In response to the absence of a literature survey in this domain, this paper offers an overview of existing in-IDE HAX research, distilling main directions and insights to guide future investigations.

Our literature survey was conducted with a focus on inclusivity and reliability. We selected prominent online databases known for their relevance and rigor in the fields of software engineering, AI, and HCI, specifically choosing the ACM Digital Library, DBLP, IEEE Digital Library, and ArXiv. The selection of ArXiv was particularly important due to the rapid development in our field. Our search was targeted at research articles published from $2020$ onwards, aligning with the recent advancements in the field. This initial search yielded a pool of $211$ papers. We then carefully applied a set of criteria to ensure only papers directly relevant to in-IDE HAX were included. This filtering process led to the selection of $36$ significant papers, published between $2020$ and $2024$, which collectively offer a comprehensive perspective for our study.

Analyzing the papers, we managed to identify three key research areas. The first group of studies~\cite{tan2023copilot, Lehmann2020Examining, mcnutt2023design, weisz2023general} focuses on integrating AI assistants into programming environments, emphasizing the impact of user interface and proposing design principles for user-friendly AI assistance. Second group~\cite{gu2023data, prather2023itsweird, weisz2022better} explores how AI assistance reshapes and impacts the programming workflow, categorizing interactions within it. The third category of studies~\cite{Nguyen2022Empirical, pearce2021asleep, dakhel2023github} assesses AI assistant functionality, examining correctness, understandability, and security. 
 Together, these studies give us important insights into the \textbf{Design, Impact, and Quality} of in-IDE HAX, guiding future research in this fast-changing area.

The recognition of design principles provides insights and indicates promising directions for future research, extending their application beyond conventional chat-based and code-completion formats. Understanding the shift in programmers' workflow emphasizes the importance of dedicated research to support users in adapting to these changes, especially in educational settings where accommodating workflow shifts significantly impacts the learning experience. Regarding quality, a clear trajectory for future investigations involves fine-tuning models to enhance security and correctness. This is particularly crucial for addressing non-trivial tasks and accommodating varied programming languages.

The present work not only consolidates the existing knowledge in the field of in-IDE HAX but also provides a foundation for future research efforts. The identified research areas, curated dataset, and proposed directions contribute to the collective understanding of the evolving dynamics between Humans and AI within IDEs.


\section{METHOD}

To understand the landscape of existing research in the field of in-IDE HAX, we conducted a literature survey with the following Research Questions: \textbf{``What research has already been done in this field'' and ``What are the dimensions of Human-AI Interaction inside IDE that should be addressed in the research aiming at supporting developers at their work''}. 

We took the following steps to ensure the inclusivity and reliability of the literature review.  

We first selected four well-known digital libraries to encompass a diverse range of scholarly sources. This includes ACM Digital Library, DBLP, IEEE Digital Library, and ArXiv. We intentionally included ArXiv in the list to account for emerging fields, where useful insights might be collected from not peer-reviewed papers.

Next, we employed a search strategy using the following search string to identify relevant research papers: 

\texttt{(``Integrated Development Environment'' OR ``IDE'' OR ``VS code'' OR ``Intellij'' OR ``vim'' OR ``Code editor'') AND (``AI assistant'' OR ``AI features'' OR ``Human-AI interaction'' OR ``Human-AI collaboration'' OR ``AI coding assistant'' OR ``LLM-based code generation'' OR ``AI code generation'' OR ``Conversational AI'')}. 

We then applied several inclusion and exclusion criteria to filter the retrieved literature. We excluded studies published before 2020, as well as blog posts. Additionally, we filtered out non-English manuscripts and those with titles, abstracts, or full texts that are thematically irrelevant ensuring the relevance and recency of the selected studies. 

During the process, we initially collected $211$ papers, with most of them ($191$) being sourced from ArXiv. 
After filtering based on the criteria, we extracted relevant information, including publication year, authorship, the study's overarching goal, research questions, methodology, and key findings from each selected source.
This process resulted in $36$ studies published in various conferences (11 studies), journals (3 studies), and finally yet only on ArXiv (22 studies).
Table~\ref{tab:stats} Provides d breakdown of the number of studies per source.

\begin{table}
    \centering
    \caption{Statistics on study sources}
    \label{tab:stats}
    \begin{tabular}{p{6.9cm}c}
        \toprule       
        \textbf{Venue} & \textbf{Count} \\
        \midrule
        ACM on Programming Languages  & 1  \\  \rowcolor{gray}
        ACM Transactions on Computer-Human Interaction & 2 \\
        Conference on Human Factors in Computing Systems &  1 \\  \rowcolor{gray}
        International Conference on Software Engineering &  2  \\
        International Conference on Mining Software \newline Repositories &  1  \\  \rowcolor{gray}
        International Conference on Evaluation and \newline Assessment in Software Engineering &  1  \\
        International Conference on Intelligent User \newline Interfaces &  2  \\  \rowcolor{gray}
        International Symposium on Machine Programming &  1  \\
        Technical Symposium on Computer Science Education &  1  \\  \rowcolor{gray}
        Symposium on User Interface Software and \newline Technology &  2  \\
        ArXiv &  22  \\
        \midrule
        Total      & 36  \\
        \bottomrule
    \end{tabular}
\end{table}

\section{FINDINGS}


Our literature survey revealed notable insights into the integration of AI assistance in IDEs. It is clear that this integration goes beyond altering the user interface; it fundamentally changes developers' workflows and the core models of AI assistants. Consequently, as in-IDE HAX evolves, we are seeing the growth of three distinct research areas, each with its specific focus:

    First is the exploration of User Interface \textbf{Design} in AI-enabled tools. This area seeks to answer key questions about design considerations when incorporating AI technologies into programming environments. Out of the $36$ papers we reviewed, $14$ specifically addressed this topic.
    
    Second is the investigation of the \textbf{Impact} of HAX. This line of research centers on understanding how users interact with, perceive, and benefit from AI-based tools in different programming scenarios. It addresses common themes like usability issues, the effect on productivity, and user trust in these tools. Of the $36$ papers we reviewed, $13$ explored this aspect of the research field.
    
    Third is the investigation of the \textbf{Quality} of AI Assistants. This research area delves into the performance of AI assistants, specifically looking at factors like correctness, understandability, security, and their capability to solve algorithmic problems. Among the $36$ papers we reviewed, $9$ papers specifically focused on examining these aspects of AI assistants.

\medskip
\textbf{Studies of Design of Integration}~\cite{tan2023copilot, Lehmann2020Examining, mcnutt2023design, weisz2023general, vaithilingam2023towards, Manfredi2023Mixed, Robe2022Designing, Jayagopal2022Exploring, Angert2023Spellburst, nam2023inide, wang2023investigating, cheng2023prompt, ross2023programmer, feng2023coprompt} focus on prototyping integration of AI assistants into existing programming environments, systematizing methodologies and proposing principles for designing AI-assisted programming.
Findings in this field provide evidence that the user interface of in-IDE AI assistance affects the usefulness of this tool and should be built thoughtfully. 

The design principles for AI assistance, as detailed in the papers~\cite{Lehmann2020Examining, mcnutt2023design, vaithilingam2023towards, weisz2023general}, highlight the importance of clear communication and user control in generative AI, while code assistants and autocompletion features aim for adaptability, clarity, and user-friendly interactions.


Generative AI design principles involve communicating probabilistic nature and variable outputs, facilitating user annotation and visualization of differences, accommodating imperfection through feedback, implementing user-driven controls, providing sandbox environments, orienting users through mental models, and communicating capabilities and limitations, with a focus on preventing risks and harms~\cite{weisz2023general}. Code assistants, following principles from~\cite{mcnutt2023design}, act as adaptable ghostwriters, providing ambient or interactive annotations, offering context control, balancing politeness and promotion, integrating search and documentation, shaping user expectations, and incorporating means of verification into code generation. Autocompletion features a user interface with suggestions near the input, allowing user decisions and leveraging partial input for predictions~\cite{Lehmann2020Examining}. Autocompletion design principles include glanceable suggestions, juxtaposition for clarity, simplicity through familiarity, sufficient visibility for validation, and snoozability of suggestions to prevent interruptions~\cite{vaithilingam2023towards}.


Multiple papers delved into presenting AI as a pair programmer to assist novices in their education~\cite{Manfredi2023Mixed,Robe2022Designing}, revealing a generally high level of acceptance and usefulness. Despite this, challenges persist in designing a user-friendly interface for such agents and need to be addressed.

The other set of papers investigates the \textbf{Design and Impact} of various AI integrations into IDEs. It delves into how diverse forms of interaction influence novices' workflows~\cite{Robe2022Designing,Manfredi2023Mixed,Jayagopal2022Exploring}, the level of user trust in the assistant~\cite{nam2023inide, wang2023investigating}, and user productivity~\cite{ross2023programmer, feng2023coprompt, cheng2023prompt, Angert2023Spellburst}. The results provided suggest that conversational and explainable AI might support user productivity and trust while, in some cases, causing overreliance.

\medskip

\textbf{Studies of the Impact of Interaction}~\cite{gu2023data, prather2023itsweird, weisz2022better, vasiliniuc2023case, kazemitabaar2023novices, saki2022isgithub, zhou2023concerns, mozannar2023reading, Barke2023Grounded, amoozadeh2023trust, vaithilingam2022expectation, liang2023largescale, Ziegler2022Productivity} aim to understand how programmers use and perceive AI assistants. The investigations also include identifying challenges faced by developers, creating a taxonomy of programmer activities involving AI, and assessing users' productivity, as well as their level of trust in these tools.

Research in this field shows that in-IDE Human-AI Interaction significantly affects and changes the developers' workflow. 

Using AI tools increases productivity but may involve a trade-off in code quality since developers sometimes struggle to receive from AI the outputs that would align with their requirements and expectations~\cite{liang2023largescale, weisz2022better, vasiliniuc2023case, vaithilingam2022expectation, saki2022isgithub, gu2023data, Ziegler2022Productivity}. The context in which AI tools are used, the quality of suggestions, and compatibility issues play crucial roles in shaping these tools' overall effectiveness and user perception.

The findings emphasize that interactions between programmers and AI assistants alter the traditional programming workflow. This transformation introduces dedicated time for interacting with AI and processing its outputs~\cite{mozannar2023reading} and can be categorized into various modes of HAX~\cite{Barke2023Grounded, prather2023itsweird}. The findings suggest that successful AI integration involves balancing the benefits of acceleration with the need for careful exploration and validation.

In assessing the impact of AI assistance on novices, it becomes evident that while AI tools can positively influence programming education by facilitating learning and boosting motivation, challenges such as over-reliance and the need for cautious acceptance require attention~\cite{prather2023itsweird, kazemitabaar2023novices, amoozadeh2023trust}. These challenges might be addressed by implementing educational strategies and awareness programs.

\medskip
\textbf{Studies of Quality of Interaction}~\cite{Nguyen2022Empirical,pearce2021asleep, dakhel2023github, Zhou2022Improving, asare2023githubs, sandoval2023lost, huang2022se, Wermelinger2023Using, mozannar2023suggestion} are qualitatively and quantitatively evaluating various aspects of AI assistants' functionality, ranging from correctness and understandability of suggested code to investigating security concerns.

These studies reveal that the effectiveness and utility of an AI assistant depend not only on its user interface but also on the quality of the model's output. Moreover, there is no one-size-fits-all solution for the core model choice, as variations can impact user productivity and the acceptance rate of suggestions.

Most of the works are focused on the correctness of LLM outputs, analyzing if the suggestions provided are relevant to the task and error-free~\cite{dakhel2023github, Nguyen2022Empirical, huang2022se, Wermelinger2023Using, sandoval2023lost, Zhou2022Improving}. Findings suggest that while AI assistants can provide relevant solutions and suggestions, especially for trivial types of tasks, they still might be erroneous and require the user's ability to notice and correct them. 

Regarding the comprehensibility and complexity of the code, several research show that the assistants generally produce understandable code that might be even less complex than one written by human~\cite{dakhel2023github, Nguyen2022Empirical, Wermelinger2023Using}.

A few studies are focused on security, assessing if the assistance introduces security vulnerabilities to the code and to what extent~\cite {sandoval2023lost, pearce2021asleep, saki2022isgithub}. From these studies, it can be derived that security risks might depend on the model used for the assistance foundation since, in some cases, the rate of vulnerability was low, but in some cases relatively high, reaching 40\% of generated programs on C language appeared to be vulnerable. 

Some papers covered how fine-tuning the foundational model can help to improve the quality of interaction~\cite{Zhou2022Improving,huang2022se, mozannar2023suggestion} --- support accuracy and timing of the shown suggestions to enhance the acceptance rate. 

Exploring the interplay between models' characteristics, user interfaces, and real-world usage can provide valuable insights into optimizing AI assistants for diverse programming tasks and improving developers' overall experience.
\section{DISCUSSION}

In examining the literature, three distinct categories emerge, each shedding light on various aspects of AI assistance. 

    The first category explores the \textbf{Design} of HAX, drawing insights from interviews, literature reviews, experiments, or experiences in other domains. Some papers within this category focus solely on design, while others extend their work to investigate how different forms of AI impact users. 
    
    The identified design principles, including a seamless user experience, visibility of suggestions, and considerations for context control, snoozability, and potential harm mitigation, form a foundation for developing user-friendly AI assistance tools. The challenges associated with presenting AI as a pair programmer for novices underscore the need for ongoing efforts in designing interfaces that cater to diverse user levels, ensuring acceptance and usefulness.

    The second category delves into the \textbf{Impact} of HAX, questioning whether its presence affects users and, if so, in what ways. 

    The integration of AI tools, while enhancing productivity, introduces a nuanced trade-off in code quality. Developers grapple with aligning AI-generated outputs with their requirements, highlighting the intricate interplay of contextual factors, suggestion quality, and compatibility issues. The identified modes of HAX and the necessity for careful exploration and validation underscore the need for a balanced approach, ensuring the positive acceleration of workflow without compromising code quality.

    The third category centers on the \textbf{Quality} of HAX, probing into the ability of AI to provide secure, comprehensible, and adequate code suggestions.  

    The findings emphasize that the effectiveness of AI assistants hinges not only on the user interface but also on the quality of the underlying models. Notably, studies highlighted the variability in the correctness of LLM outputs, indicating a need for user vigilance. Comprehensibility and code complexity were identified as areas where AI assistants generally perform well. However, security assessments revealed potential vulnerabilities, underscoring the importance of fine-tuning foundational models to enhance overall interaction quality.

\subsection{FUTURE WORK}

Specifically, within our area of interest, we suggest focusing on three aspects: \textbf{Task-specific user interface} in the realm of Design of Interaction, \textbf{Trust} in Impact of Interaction, and \textbf{Readability} in Quality of Interaction. 

In the domain of the \textbf{Design} of HAX, we propose that chat-based interaction with LLMs may not always be the most effective approach, and different tasks may require alternative methods. We suggest two complementary directions within this area: (1) Analyzing common challenges addressed through chat-based models and implementing them in more convenient non-chat formats. For instance, writing method documentation is efficiently accomplished through IDE plugins with context menu buttons.~\footnote{Documentation generation is a feature in AI-assistant plugin for IntelliJ IDEA: https://plugins.jetbrains.com/plugin/22282-ai-assistant} 
(2) Enhancing model reactivity by transforming predictable actions into automatic suggestions presented to the user. For example, automatically proposing fixes after an error or suggesting the optimization after using the profiler. This enables a seamless and proactive AI interaction experience. While we believe chat-based interaction creates a lot of traction in Human-AI Interaction, it is an interesting challenge to try to improve inherently chat-based interactions --- like question answering about the code base. Lastly, there is uncharted territory of the interfaces for the no-code and multi-agent development tools. What shape will the development process be if users interact with several AI assistants or even create software using only natural language without code itself? 

By focusing on trust, we aim to account for the developers' attitudes to AI, which might \textbf{Impact} the interaction with technology. One possible approach to the trust problem is looking at the Natural Language Processing domain. Such approaches as highlighting the tokens that affected the output the most~\cite{NIPS2017_7062}, approximating the uncertainty of the model~\cite{huang2023look} or just providing clear and transparent context which was provided to the model could be beneficial for facilitating trust between AI and the user. While such methods have already been developed, it is still a challenge to introduce them into the context of the IDE without hurting user productivity and the development process. 

Regarding the \textbf{Quality} of interaction, we believe that readability is a promising concept for code models' alignment --- we suggest that given equal correctness of different models' outputs, people would prefer the model that will produce more readable code since such outputs will improve users' productivity while working with it. While there are multiple methods of LLMs alignment developed~\cite{christiano2017deep}, there are just a few models of code readability~\cite{scalabrino2018comprehensive} and studies about how people discriminate between readable and non-readable code~\cite{scalabrino2017automatically}. Therefore, it is an open question whether we can align a model to consistently produce more code that will be perceived as more readable by actual users. In this direction of work, we suggest looking for general properties of code that people find desirable to see, or at least some strata of programmers with a common view of what desirable code is, and providing for the specific fine-tuning.

\subsection{THREATS TO VALIDITY}
While gathered findings provide valuable insights, in any scientific study, it is essential to consider potential threats to validity, which are factors that can affect the accuracy and generalizability of the research findings. 

\textbf{Sampling Bias:} Despite rigorous efforts to include well-known libraries and refine the search string, the possibility of sampling bias remains. While our approach aimed to minimize bias, the inherent challenge of capturing every relevant work persists. That is why we provide our search protocol to mitigate this threat.

\textbf{Temporal Bias:} The chosen timeframe (papers published between 2020 and 2024) introduces a potential temporal bias, excluding earlier works. This decision was driven by the intention to focus on contemporary developments following the advent of LLMs. While acknowledging potential temporal limitations, we aimed to capture the latest advancements and trends in this rapidly evolving field.

\textbf{Source Reliability:} Inclusion of non-peer-reviewed papers from ArXiv introduces concerns regarding the reliability of findings, given the absence of a formal peer-review process. Recognizing this limitation, we deemed it necessary to consider insights from ArXiv due to the dynamic and rapidly evolving nature of the field. Thoroughly investigating titles, abstracts, and full texts ensured that all included papers contributed meaningfully to our survey. Moreover, the open publication environment of ArXiv encourages publishing negative or null results, contributing to a more balanced representation of research outcomes.

\textbf{Interpretation Bias:} Analyzing a large amount of information may introduce interpretation bias and impact how studies are categorized. While acknowledging this complexity, we emphasize transparency and provide the entire dataset.~\footnote{Dataset: https://doi.org/10.5281/zenodo.10290921} for readers

\section{CONCLUSION}

The field of in-IDE Human-AI Experience has just started to develop, but some patterns and overall conclusions might be derived from the published works already. It is evident that the field has three branches, namely \textbf{Design, Impact, Quality} of Interaction. In the dimension of Design, research provide design principles for creating in-IDE AI assistance and making them usable, as well as exploring efficient options for AI integration into the educational process and IDEs of various kinds. The dimension of Impact mostly focuses on changes in developers' workflow and efficiency, as well as their overall experience with AI assistants. The dimension of Quality studies if the foundational models of AI assistants are correct enough, producing comprehensible and secure code, and if this quality can be increased by fine-tuning practices. Based on this discovery, we propose several specific research topics in each of the dimensions that might be covered by industrial research teams to address open questions in the field.


\bibliographystyle{ACM-Reference-Format}
\bibliography{main_refs}


\end{document}